\def\decsec{\hbox{$^{\prime\prime}\hskip-3pt .$}}

\def\hii{H~{\sc II}}

\documentclass{kapproc} 

\usepackage{procps} 

\usepackage[dvips]{graphicx}

\upperandlowercase

\kluwerbib 

\begin{document}



\articletitle{Are there local analogs of Lyman break galaxies?}















%


\author{James D. Lowenthal\altaffilmark{1}, R. Nick Durham\altaffilmark{1}, 
         Brian J. Lyons\altaffilmark{2}, Matthew
         A. Bershady\altaffilmark{3}, Jesus Gallego\altaffilmark{4},
         Rafael Guzman\altaffilmark{5}, David C. Koo\altaffilmark{6}} 

\affil{\altaffilmark{1}Smith College, Northampton, MA, USA 01060 \
\altaffilmark{2}Amherst College \
\altaffilmark{3}University of Wisconsin \
\altaffilmark{4}Universidad Complutense de Madrid \
\altaffilmark{5}University of Florida \
\altaffilmark{6}UCO/Lick Observatory}




 \begin{abstract}
 To make direct comparisons in the rest-far-ultraviolet between LBGs at
$z\sim3$ and more local star-forming galaxies, we use HST/STIS to
image a set of 12 nearby ($z<0.05$) \hii\ galaxies in the FUV and a
set of 14 luminous compact blue  galaxies (LCBGs) at moderate redshift
($z\sim0.5$) in the NUV, corresponding to the rest-FUV.  We then
subject both sets of galaxy images and those of LBGs at $z\sim3$ to
the same morphological and structural analysis.  We find many
qualitative and quantitative similarities between the rest-FUV
characteristics of distant LBGs and of the more nearby starburst
samples, including general morphologies, sizes, asymmetries, and
concentrations.  Along with some kinematic similarities, this implies
that nearby \hii\ galaxies and LCBGs may be reasonable local analogs
of distant Lyman break galaxies.

 \end{abstract}



\section{Lyman break galaxies}
Lyman break galaxies (LBGs) at redshifts $z\sim3$ are the current gold
standard for star-forming galaxies in the early universe, at least
rest-UV and -optically selected ones.  Many of their properties are
revealed by deep multi-wavelength imaging and spectroscopic surveys
(e.g., presentations at this conference by Bremmer, Erb, Huang,
Mehlert, Papovich, Sawicki, and others).  These include small sizes
$r_{1/2}<4$ kpc, high luminosities $L\sim L*$, significant clustering, and
diverse morphologies.  LBGs are also copiously star-forming, easily
qualifying as starbursts according to the star formation intensity
(SFR per unit mass or gas mass) definition advocated by Tim Heckman at
this conference (e.g., Meurer et al. 1997).

Significant questions remaining about the nature of LBGs include their
masses, their mass assembly histories, their fate, and their
environments, including any dark matter.  Comparison to local galaxies
with similar properties may help illuminate some or all of those
issues, since nearby systems can generally be studied in much greater
detail.

\section{Compact starbursts at $z<1$ in the Rest-UV}

We face two problems in attempting to draw parallels between distant
LBGs and nearby starbursts: (1) LBGs are best seen in the optical,
corresponding at $z=3$ to the rest-UV; and (2) it is not obvious which
kind or kinds of local systems are the best proxies.  

Two classes of galaxy at $z<1$ seem especially promising as nearby
cousins of LBGs: \hii galaxies at $z\sim0$ and luminous compact blue
galaxies (LCBGs) at $0.4<z<1$.  Both classes show, in the optical, the
small sizes, high luminosities, diverse morphologies, and copious star
formation that also characterize LBGs.  To push the comparison further, we
have obtained rest-UV images with the Space Telescope Imaging
Spectrograph onboard  {\it Hubble Space Telescope} (HST/STIS) of 12
\hii\ galaxies and 14 LCBGs.  The samples are drawn from the UCM
survey (Perez-Gonzalez et al. 2001) and the Kitt Peak Galaxy Redshift Survey (Munn 1997),
respectively.  Many have also been imaged with HST/WFPC2 and/or
NICMOS and/or studied spectroscopically at Keck and Arecibo
(e.g. Pisano et al. 2001).

\section{Rest-UV Morphologies}

\begin{figure}
\vspace{7.5cm}
\includegraphics{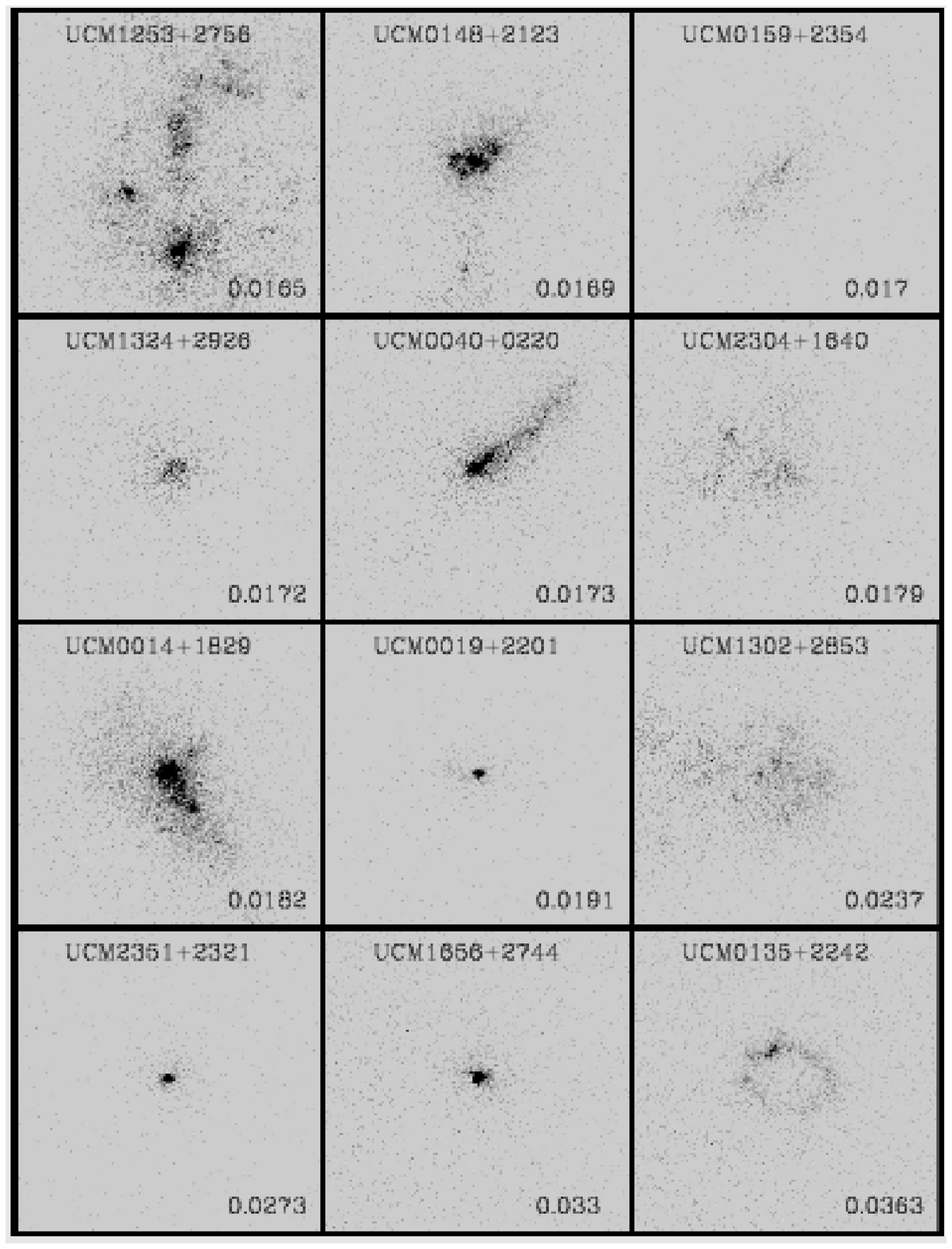} 
\includegraphics{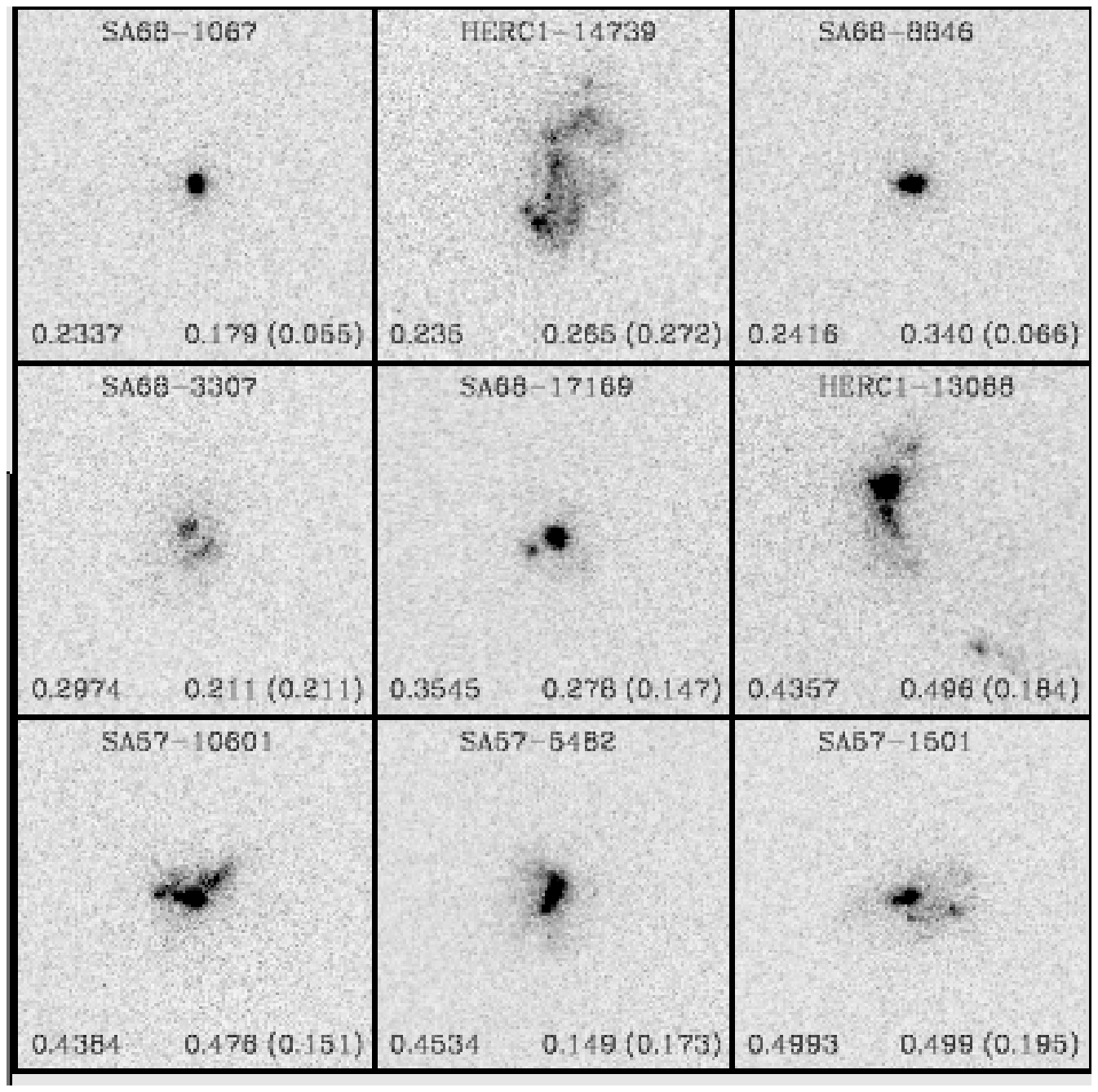} 
\caption{\label{mosaic} Left: Mosaic of HST/STIS-FUV images
($\lambda_c=1590$\AA) of 12 \hii\ galaxies from the UCM survey.
Redshifts are labelled for each galaxy.  Each image is 6\decsec5 on a
side, corrsponding to 2.5 kpc at $z=0.02$.  Right: Mosaic of
HST/STIS-NUV images ($\lambda_c=2320$\AA) of 9 of the 14 LCBGs.  Each
image is 3\decsec75 on a side, corresponding to 22.5 kpc at
$z=0.5$.  Note the small sizes and disturbed, varied rest-UV
morphologies of both samples, reminiscent of LBGs at higher redshift.}
\end{figure}

The HST/STIS UV images of the 26 low- and intermediate-redshift
starburst galaxies in our sample are shown in Fig.~\ref{mosaic}.  It
is immediately obvious that the rest-UV morphologies represent a
diverse panoply, rather than a uniform class.  Multiple knots, tails,
and extended emission -- almost all invisible at ground-based
resolution in the optical -- are the rule rather than the exception.

\begin{figure}
\vspace{4cm}
\includegraphics{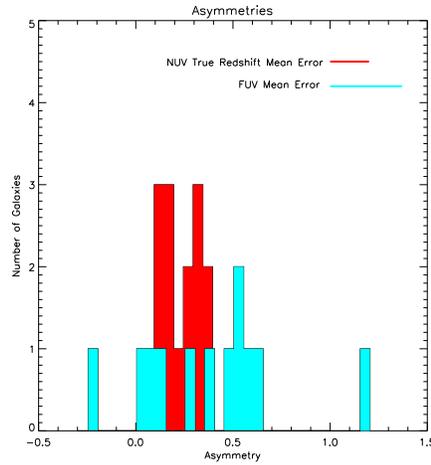} 
\caption{\label{asym} Distribution of asymmetries measured at
Petrosian radius for both the \hii\ galaxy (FUV) and LCBG (NUV)
samples.  The mean is roughly consistent with that of LBGs measured in
the HDF by Conselice et al. (2003).}
\end{figure}

We attempted to quantify the UV morphologies of our nearby and
intermediate compact starbursts using the CAS (compactness, asymmetry,
and clumpiness) methodology of Conselice et al. (2003).  The
measured asymmetry is very sensitive to the exact radius at which it
is measured; we tried both half-light radii and Petrosian radii.
Fig.~\ref{asym} shows the distribution of asymmetries $A(r_P)$.  We
find that the mean $A(r_P)$ of our two samples is roughly consistent
with that of LBGs measured in the HDF by Conselice et al. (2003),
although the formal measurement uncertainties are very large.

\section{Simulating LBGs at $z=3$}

\begin{figure}
\vspace{7cm}
\includegraphics{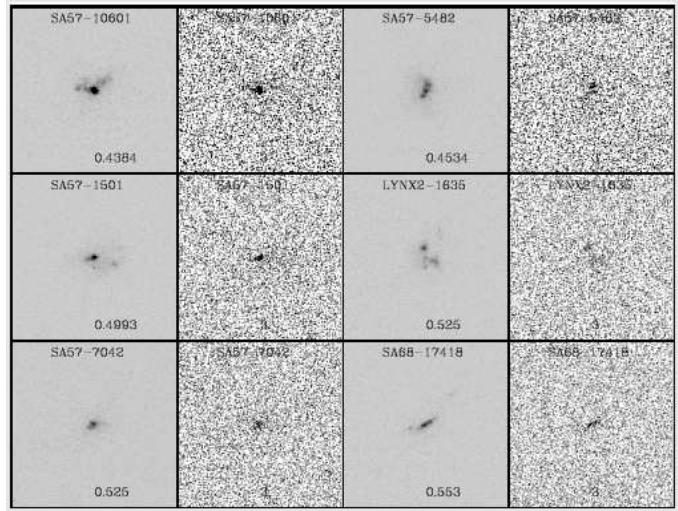} 
\caption{\label{noisemosaic} LCBGs shown at their true redshift (left
panel of each pair) and simulated view as they would appear at
redshift $z=3$ observed in the HDF.  All sources would be easily
detected at $z=3$, although faint, low-surface brightness emission is
lost to noise.}
\end{figure}

What would the \hii\ galaxies and LCBGs look like if placed at
redshift $z=3$?  We simulated that view by resampling our HST/STIS
images and adding noise to reproduce the Hubble Deep Field
sensitivity.  The \hii\ galaxies are too faint to detect, but the
LCBGs, which are more luminous, are all easily detected
(Fig.~\ref{noisemosaic}).  Despite the loss of low-surface brightness
features, the similarities to the appearances of real LBGs are
striking.

We conclude that \hii\ galaxy and LCBG morphologies, like their sizes,
colors, and star-formation rates and intensities, are qualitatively
and quantitatively similar to those of LBGs.  They are therefore
reasonable local testbeds for further comparative study of LBGs,
including constraints on mass.

Our future plans include applying other morphological measures such as
the Gini coefficient (Lotz et al. 2004) and combining these STIS UV
images with WFPC2 and NICMOS images in hand to constrain stellar
populations, dust content, and merger scenarios.






%





\begin{chapthebibliography}{<widest bib entry>}


\bibitem{cons03} Conselice, C.J., Bershady, M.A., Dickinson, M. \& Papovich,
C. 2003, ApJ, 136, 1183

\bibitem{koo95}  Koo, D.C., Guzman, R. Faber, S.M., Illingworth,
G.D., Bershady, M.A., Kron, R.G., \& Takamiya, M. 1995 ApJL, 440, 49 

\bibitem{guzma98} Guzman, R., Jangren, A. Koo, D.C. Bershady, M.A., \&
Simard, L. 1998, ApJL, 495, 13

\bibitem {lotz04} Lotz, J. M., Primack, J., \& Madau, P.  2004, AJ, 128, 163

\bibitem {meur97} Meurer, G., Heckman, T., Lehnert, M., Leitherer, C. \&
Lowenthal, J.D. 1997, AJ, 114, 54.  

\bibitem {munn97} Munn, J.A., Koo, D.C., Kron, R.G., Majewski, S.R., Bershady, M.A., \& Smetanka, J.J. 1997, ApJS, 109, 45

\bibitem {pere01} Perez-Gonzalez, P.G., Gallego, J., Zamorano, J., \&
Gil de Paz, A. 2001, A\&A, 365, 370

\bibitem {pisa01} Pisano, D. J., Kobulnicky, H. A., Guzman, R., Gallego, J.,
\& Bershady, M.A. 2001, AJ, 122 1194 

\end{chapthebibliography}
\end{document}